# Ballistic and resonant negative photocurrents in semiconducting carbon nanotubes

Christoph Karnetzky,[1,2] Lukas Sponfeldner,[1] Max Engl,[1] and Alexander W. Holleitner[*1,2]

[1]*Walter Schottky Institute and Physics Department, Technical University Munich, Am Coulombwall 4(a), 85748 Garching, Germany.*

[2]*Nanosystems Initiative Munich (NIM), Schellingstr. 4, 80799 Munich, Germany.*

Ultrafast photocurrent experiments are performed on semiconducting, single-walled carbon nanotubes under a resonant optical excitation of their subbands. The photogenerated excitons are dissociated at large electric fields and the resulting transport of the charge carriers turns out to be ballistic. Thermionic emission processes to the contacts dominate the photocurrent. The charge current without laser excitation is well described by a Fowler-Nordheim tunneling. The time-averaged photocurrent changes polarity as soon as sufficient charge carriers are injected from the contacts, which can be explained by an effective population inversion in the optically pumped subbands.

Keywords: carbon nanotubes, exciton dissociation, optoelectronic dynamics, non-equilibrium 1D transport, ballistic transport

PACS numbers: 85.60.-q, 61.48.De, 73.23.Ad

Carbon nanotubes (CNTs) exhibit one-dimensional electron systems with a large exciton binding energy of several hundreds of meV[1,2] and they allow for the fundamental investigation of exciton- and electron-phonon interactions in reduced dimensions.[3-5] Despite the large exciton binding energy, the generation of a photocurrent has been reported in several experiments.[6,7] It is largely discussed in terms of photo-thermoelectric effects in combination with electric fields induced by potential fluctuations and contact potentials.[7-14] However, there is still very little known about the temporal dynamics of the charge transport to the contacting reservoirs with regard to the exciton dissociation as well as the relaxation and recombination dynamics within the CNTs' subbands. Recent work on p-i-n junctions in CNTs reveals a diffusive transport of photogenerated charge carriers to the contacts with an onset of ballistic transport at high electric fields.[15] For unbiased CNTs, further work introduces a 'spontaneous dissociation' of excitons giving rise to a photocurrent, without clarifying the underlying transport processes.[16] Moreover, it is unclear whether the mechanism of the photo-induced non-equilibrium charge transport can be distinct to the charge transport without laser excitation.

We reveal the ultrafast, non-equilibrium transport properties of photogenerated electrons and holes in few to single semiconducting CNTs. We compare the results for CNTs where either the second or first subband is resonantly excited. We demonstrate that the ultrafast photocurrent in the CNTs is dominated by a ballistic transport. By a time-of-flight analysis, we resolve a ballistic group velocity of the photogenerated charge carriers. Moreover, we identify a thermionic emission of the photogenerated charge carriers to the contacts. In this picture, photogenerated charge carriers with a high kinetic energy can overcome the energy barriers to the contacts, and they drive the overall photocurrent. The mechanism stands in contrast to the process which we find for the so-called dark current, i.e. the charge transport without laser excitation. Here, a Fowler-Nordheim tunneling of charge carriers from the contacts to the CNTs consistently describes the data. The overall slowest optoelectronic processes occur on a nanosecond time-scale. We detect them at very high bias voltages, and they are consistent with a so-called lifetime-limited photocurrent, as recently reported for ensembles of CNTs.[17] In time-averaged measurements, we observe a sign change of the photocurrent for a high bias, which we explain by an effective population inversion of the optically pumped subband of the CNTs via charge tunneling processes from the metal contacts. Our experiments give fundamental insights into the ultrafast dynamics of photogenerated charge carriers in contacted, semiconducting CNTs ranging from the photocurrent generation to the non-equilibrium transport of the charges to and from the contacts. The insights may prove essential for ultrafast optoelectronic devices and photodetectors based on semiconducting CNTs in general, but particularly, on single CNTs integrated into optoelectronic high-speed circuits and THz-striplines.

The experiments are performed on two sets of semiconducting CNTs. The first is synthesized by the arc-discharge method and the second by the CoMoCat synthesis.[18,19,20] The first (second) CNTs have a diameter of $d_{CNTS}$ ~ 1.5 nm (0.8 nm). Via dielectrophoresis,[21] the CNTs are deposited in-between two Ti/Au contacts with a height of 10/300 nm [Fig. 1(a)]. The contacts are fabricated by optical lithography and they form lateral, coplanar striplines with a total length of ~58 mm, a width of 5 μm, and a separation of 10 μm. The striplines are utilized to perform the ultrafast, time-resolved photocurrent experiments.[22-25] At the position of the CNTs, the distance between the striplines is reduced to (1.1 ± 0.1) μm. All measurements are performed at ~$10^{-6}$ mbar and 77 Kelvin in a cryostat. We use a fiber-based pulsed laser with a pulse duration of <30 fs, a photon energy continuum between 0.9 and 1.3 eV, and a repetition frequency of 80 MHz. For measuring the time-integrated photocurrent $I_{photo}$, the photon energy $E_{photon}$ is further filtered by using a monochromator such that the laser power $P_{laser}$ amounts to ~100 W/cm$^2$ on the CNTs per center wavelength. We confirm the positioning of few to single CNTs between the contacts by using a scanning photocurrent microscopy with a lateral resolution of about 2 μm [Fig. 1(b)]. Fig. 1(c) shows the photocurrent spectrum of $I_{photo}$ vs. $E_{photon}$ measured at the position in-between the two metal contacts. For all applied bias voltages $V_{sd}$, the photocurrent exhibits a clear maximum at $E_{photon}$ ~ 1.15 eV. This energy coincidences with the anticipated transition energy $E_{22} = C_2 - V_2$ between the second conduction (valence) band $C_2$ ($V_2$) for the given diameter $d_{CNTS}$ ~ 1.5 nm of the CNTs.[26] We explain the relatively broad full width at half maximum $FWHM = (107 \pm 6)$ meV by intra-CNT charge carrier dynamics or by different species of CNTs.[21] Fig. 1(c) already demonstrates that a photocurrent is measured even at zero bias and for voltages much smaller than the exciton binding energy.

Fig. 2(a) shows a false-color plot of the time-integrated photocurrent $I_{photo}$ vs. $E_{photon}$ and $V_{sd}$. For $|V_{sd}| \geq 8$ V, we observe a sign change of $I_{photo}$ which is clearly seen in the line scans $I_{photo}$ vs. $V_{sd}$ for a fixed photon energy [$E_{photon} = 1.15$ eV in Fig. 2(b)]. The dashed line in Fig. 2(b) describes a thermionic emission process with barrier lowering, i.e. Schottky emission of photogenerated charge carriers from the CNTs across the contact barriers into the metal contacts.[20] Interestingly, also in the high voltage regime where $I_{photo}$ changes sign, the amplitude of $I_{photo}$ follows this thermionic model. For comparison, Fig. 2(c) shows the dark current $I_{dc}$ without laser illumination for the same bias regime. We observe that $I_{dc}$ does not show a sign change and that the data can be fitted by a Fowler-Nordheim tunneling [dotted line in Fig. 2(c)]. Considering the voltage dependences of $I_{photo}$ and $I_{dc}$, we identify three voltage regimes I, II, and III for CNTs, in which the second subband is resonantly excited. In regime I, $|V_{sd}| \leq 2.5$ V, no $I_{dc}$ passes through the sample, while a laser excitation leads to photogenerated charge carriers and hence to a finite $I_{photo}$. In regime II, $2.5$ V $\leq |V_{sd}| \leq 8$ V, a finite $I_{dc}$ can be measured in addition to $I_{photo}$. In regime III, $|V_{sd}| \geq 8$V, $I_{photo}$ changes sign while the sign change does not occur for $I_{dc}$. We note that for regime III, the amplitude of $I_{photo}$ is about three orders of magnitude smaller than the amplitude of $I_{dc}$. In this respect, the measured $I_{photo}$ can be understood as a small modulation of $I_{dc}$ by the laser excitation.

Fig. 3 sketches the band structure of the contacted CNTs excited at the second subbands. For clarity, we discuss only the dynamics for electrons, since the hole dynamics are symmetric in energy. In regime I and no laser excitation [Fig. 3(a)], we can neglect $I_{dc}$ as the thermal activation energy $k_BT \sim 7$ meV is too small for the electrons to be emitted from the metal contacts into the conduction subbands $C_1$ ($C_2$) and from the valence subbands $V_1$ ($V_2$) to the contacts. In regime I with a laser excitation resonant to $E_{22}$, electrons are optically excited from $V_2$ to $C_2$ [upward arrow in Fig. 3(b)]. Then, they relax and recombine within the CNTs (dotted and black downward arrows) and/or they propagate to the contacts (arrow to the right). The corresponding overall photocurrent can be described as $I_{photo} = e \cdot v \cdot n_e^{photo} \cdot V_{sd}$ (1), with $n_e^{photo}$ an effective density of photogenerated electrons in the CNTs, $v$ their average drift velocity, and $e$ the elementary charge. We observe a finite $I_{photo}$ at no $I_{dc}$ in this regime, which further suggests that the electron transfer from the subbands to the contacts occurs on similar timescale as the relaxation and recombination processes within the CNTs, i.e. on femto- to picoseconds.[27] This non-equilibrium scenario explains that $I_{photo}$ can be described by a thermionic emission model as expected for photogenerated charge carriers with a high kinetic energy. In regime II [Fig. 3(c)], electrons from the metal contacts can tunnel into the first subband of the CNTs generating the current $I_{dc}$ as detected in our measurement without laser excitation [Fig. 2(c)]. Consistently, we can describe $I_{dc}$ by a Fowler-Nordheim tunneling process, which sets in at a finite bias voltage.[20] We note that this current is not detected in the signal $I_{photo}$ [Fig. 2(b)], since it is not coherent with respect to the chopper reference. $I_{photo}$ is still well-described by the thermionic model in this regime. Furthermore, electrons in $V_1$ can tunnel into the drain contact, such that this subband can be assumed to have empty states available [Fig. 3(c)]. In regime III [Fig. 3(d)], electrons from the metal contacts can now also tunnel into $C_2$, where they can interact with the laser excitation. Moreover, electrons both in $V_1$ and $V_2$ can quickly tunnel to the drain contact, such that in average, both subbands have free electron states available. Overall, this scenario leads to a population inversion. Because of the conduction and valence band symmetry in CNTs, the rate for an optical transition from $C_2$ to $V_2$ equals the one from $V_2$ to $C_2$ which can be assumed to be on a sub-picosecond timescale.[28] Hence, the optical transition $C_2$ to $V_2$ occurs on a timescale comparable to the timescales of internal relaxation and recombination processes.[27] This stimulated emission then reduces the overall electron density $n_e$ in $C_2$ by an amount of $-n_e^{photo}$. The corresponding reduction of $I_{dc}$ has the opposite sign to $I_{dc}$, and it is coherent to the chopper reference. Therefore, it shows up in the signal $I_{photo}$ with a negative sign. In terms of equation (1), $I_{photo}$ can be written as $-e \cdot v \cdot n_e^{photo} \cdot V_{sd}$. Hereby, we explain the sign change of $I_{photo}$ in regime III [Fig. 2(b)]. The arguments equally apply to the hole states in the CNTs because of the mentioned electron-hole symmetry.

The discussed sequence of regimes I, II, and III is generic to semiconducting CNTs which are resonantly excited in the second subbands. We find an equivalent sign change of the photocurrent, when we resonantly pump the first subbands. This experiment has been performed on the second set of CNTs with $d_{CNTS} \sim 0.8$ nm.[20] We note that in all cases, the values of the applied bias voltages for the different regimes are determined mainly by the contact morphology. They do not necessarily correspond to the subband energy spacings within the CNTs.[29]

In order to resolve the underlying non-equilibrium within the CNTs, we perform time-resolved ultrafast photocurrent measurements (again shown for the first set of CNTs with $d_{CNTS}$ ~ 1.5 nm). We use an on-chip THz-time domain photocurrent spectroscopy where the femtosecond pump laser excites the electronic states within the CNTs. This laser is the same as for $I_{photo}$ in Fig. 2. Since the contacts form striplines, the photocurrent gives also rise to electromagnetic transients in the metal striplines with a bandwidth of up to 2 THz.[22-25] The transients run along the striplines, and they are detected on-chip by a time-delayed optical femtosecond probe pulse in combination with an Auston-switch.[22]. We use ion-implanted amorphous silicon for this ultrafast photodetector with a sub-picosecond time-resolution.[23,24] The current $I_{sampling}$ across the Auston-switch is sampled as a function of the time-delay $\Delta t$ between the two laser pulses, and it is directly proportional to the ultrafast photocurrents in the CNTs.[17]

Fig. 4(a) shows $I_{sampling}$ vs. $\Delta t$ for varying $V_{sd}$. We fit $I_{sampling}(\Delta t)$ with two functions. The first is a Gaussian $f_G$ having a $FWHM = (460 \pm 10)$ fs (dotted line). It describes the ultrafast displacement current at the CNT-Au contact of the stripline circuit,[17] which defines the moment of time when the laser pulse impinges onto the striplines with respect to the propagation of the photogenerated charge carriers in the CNTs. The $FWHM$ of $f_G$ is determined by the effective dispersion and attenuation of the THz-circuit. The second fit component is a Gaussian convoluted decay-function $f_{exp}$ with a decay time $\tau_1$ (dashed lines). Figs. 4(b) and 4(c) depict the area $A(f_{exp})$ vs. $V_{sd}$ in a linear and logarithmic scale, respectively. Again, $A(f_{exp})$ can be consistently fitted by a thermionic emission process [black lines in Figs. 4(b) and 4(c) and compare dashed lines Fig. 2(b)].[20] In our interpretation, these are the photogenerated charge carriers with an energy high enough to overcome the contact barriers, which also dominate the time-averaged signal $I_{photo}$ [compare Fig. 2(b)]. Interestingly, the relative time delay $\delta t$ of the photocurrent within the CNTs ($f_{exp}$) with respect to the displacement peak ($f_G$) varies for positive and negative $V_{sd}$ [Fig. 4(d)]. In particular, for a positive $V_{sd}$ approaching zero bias, $\delta t$ reaches $(0.8 \pm 0.1)$ ps. For a negative $V_{sd}$ approaching zero bias, $\delta t$ reaches zero. This asymmetry suggests that the photocurrents are generated at the two contacts. With a distance of $(1.1 \pm 0.1)$ μm between the two contacts, we compute a velocity $(1.1-1.6) \cdot 10^6$ ms$^{-1}$. Within the errors, this velocity agrees well with the ballistic group-velocity of the CNTs,[30] and it is significantly below a plasmon velocity.[31,32] The interpretation of a ballistic transport is further corroborated by the variation of $\delta t$ for large $V_{sd}$ [Fig. 4(d)]. There, for both polarities of $V_{sd}$, $\delta t$ reaches values of (0.4 - 0.5) ps (gray area). We argue that at large bias, the voltage drops across the whole length of the CNTs while the contact barriers become transparent, and in turn, excitons can be dissociated along the center part of the CNTs as well. Since the laser spot exceeds the length of the CNTs, $I_{sampling}$ senses the center motion of all photogenerated charge carriers along half the length of the CNT in average. Accordingly, we calculate an average propagation velocity of ½ · (1.1 ± 0.1) µm / (0.4 - 0.5) ps = (1.0 - 1.5) · $10^6$ ms$^{-1}$, which again agrees well with an anticipated ballistic group velocity. The interpretation of a ballistic propagation is substantiated by three facts. First, the deduced velocities exceed typical values for a drift saturation velocity of 5 · $10^5$ ms$^{-1}$.[3] Second, we detect a similar propagation velocity for the second set of CNTs where only the first subband is excited and solely participating in the non-equilibrium transport.[20] This is in agreement with recent estimations on the ballistic group velocity based on zone folding arguments.[30] Third, the effective mass in the first subband is about two times smaller than in the second subband. A diffusive transport of single-particle excitations would give a correspondingly shorter escape time and therefore, an increased apparent transport velocity.[15] Although we cannot perform a comparative study on the velocities in the first and second subband within one kind of CNTs (because of laser limitations), the deduced values at zero and finite bias are already at the upper limit of a ballistic group-velocity in CNTs for both sets of CNTs. Moreover, our findings at zero and large bias in Fig. 4(d) clearly indicate that the photogenerated excitons are dissociated at high electric fields, i.e. at contact potentials and at an applied bias. We propose that the dissociation of the resonantly photogenerated excitons in both the first and second subbands releases enough energy such that the electrons and holes propagate ballistically at a large kinetic energy. These are the charge carriers which show up as a thermionic current [Figs. 2(b) and 4(c)]. In principle, the length of (1.1 ± 0.1) µm exceeds the typical scattering length of high-energy optical phonons of about 100 nm.[33] In our interpretation, the peak value of $I_{sampling}$ comprises the fasted (ballistic) ensembles of photogenerated charge carriers. Slower processes are included within the decay of the time-resolved signal. For large biases, this decay time $\tau_1$ reaches values of (1.1 - 1.3) ps, while it is indistinguishable from the *FWHM* of $f_G$ close to zero bias. Generally, the photogenerated charge carriers relax within the CNTs from $C_2$ to $C_1$, for instance, via TO-phonons within a few hundreds of femtoseconds.[27] Hereby, the electron bath in $C_1$ heats up. The cooling of the electron bath is reported to occur via LA phonons on a picosecond timescale.[27] Along this line, we interpret the decay time $\tau_1$ of $f_{exp}$ to represent the relaxation and cooling of the electron bath of the CNTs in combination with a slower diffusive/drift transport regime.

We note that we do not observe a sign change of $f_{exp}$ for the highest $V_{sd}$ as we do for the time-integrated [regime III in Fig. 2(b)]. This can be explained by the utilized ultrafast measurement technique. In the THz-time domain photocurrent spectroscopy, we do not measure the charge current directly. Instead, we probe the photo-induced electric field change. Due to the symmetric band structure of CNTs, photogenerated electrons and holes induce the same electric field. Hereby, the sign change does not show up in the $I_{sampling}$. Finally, we point out that at high bias voltages, an additional decay time with $\tau_2 = (1.2 \pm 0.7)$ ns shows up.[20] We interpret the latter as the lifetime of the photogenerated charge carriers, as has been demonstrated for ensembles of CNTs.[17]

To conclude, we reveal the photocurrent generation and dynamics in semiconducting carbon nanotubes which are resonantly excited by a laser field. We find clear evidence that the photogenerated excitons are dissociated by electric fields at the contacts and within the CNTs at high biases. In a time-of-flight analysis, we extract a ballistic group velocity of the fasted photogenerated charge carriers. The dark current without laser excitation can be described by a Fowler-Nordheim tunneling. Moreover, we find that the photocurrent changes polarity as soon as the resonantly pumped subband is populated by charge carriers from the contact.

We thank the European Research Council (ERC) for financial support via project 'NanoREAL' (No. 306754).

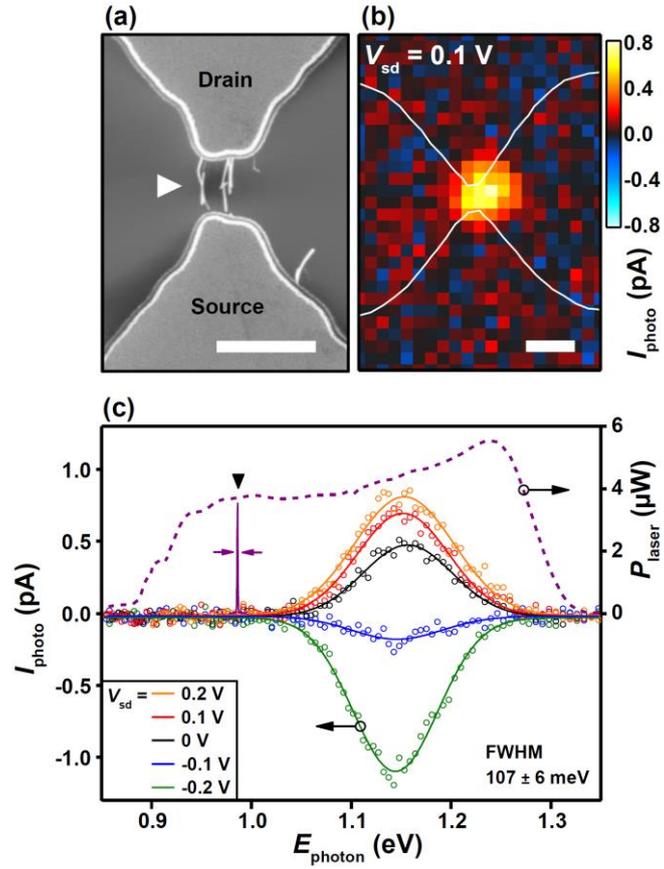

FIG 1. (color online) (a) Scanning electron microscopy (SEM) image of metal contacts and single CNTs with $d_{CNTS}$ ~ 1.5 nm (white triangle). (b) Time-integrated photocurrent $I_{photo}$ vs. spatial coordinates at $E_{photon}$ = 1.15 eV. White lines indicate edges of metal contacts. Scale bars are 2 µm. (c) Photocurrent $I_{photo}$ vs. $E_{photon}$ at the position of the maximum in (b). Dashed line shows excitation spectrum of the broadband laser. Bold triangle highlights the scanned energy window of the used monochromator.

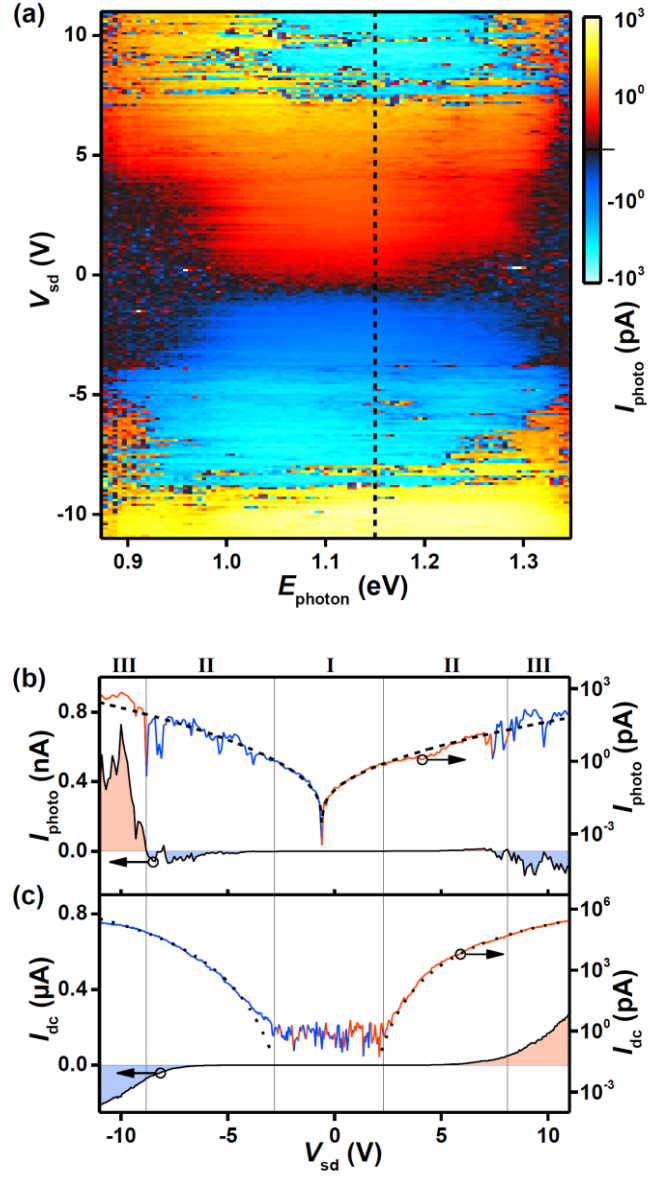

FIG 2. (a) $I_{photo}$ vs. $E_{photon}$ and $V_{sd}$ as a logarithmic color-plot for CNTs with $d_{CNTS} \sim 1.5$ nm. (b) Cross section along dashed line at $E_{photon} = 1.15$ eV in (a) in linear and logarithmic scale. Red (blue) color indicates positive (negative) sign of $I_{photo}$. (c) Dark current $I_{dc}$ vs. $V_{sd}$ without laser excitation in linear and logarithmic scale. Dashed and dotted lines are fits to the data. The regimes I, II, and III are defined in the text.

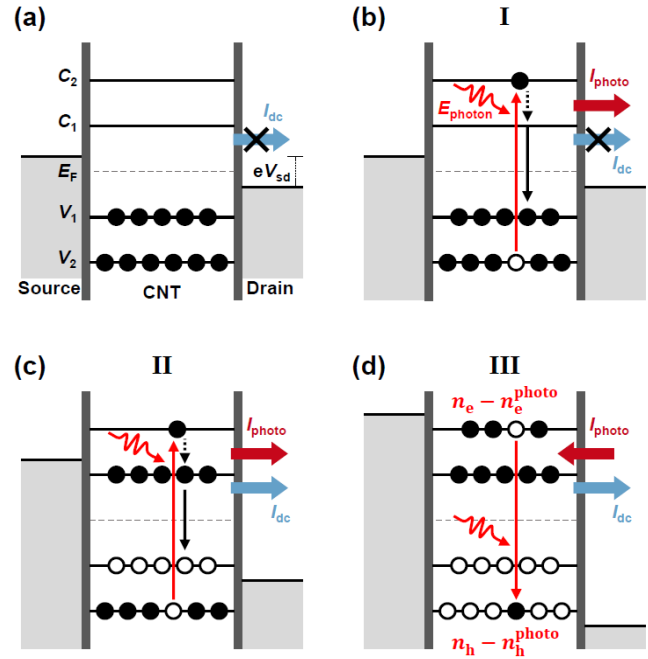

FIG 3. (color online) Sketched band diagram of CNTs for three voltage regimes I, II, and III. For simplicity, contact barriers are schematically depicted as thick vertical lines. (a) $V_{sd}$ in regime I without laser excitation. (b) – (d) $V_{sd}$ in regime I - III including the laser excitation. See text for details.

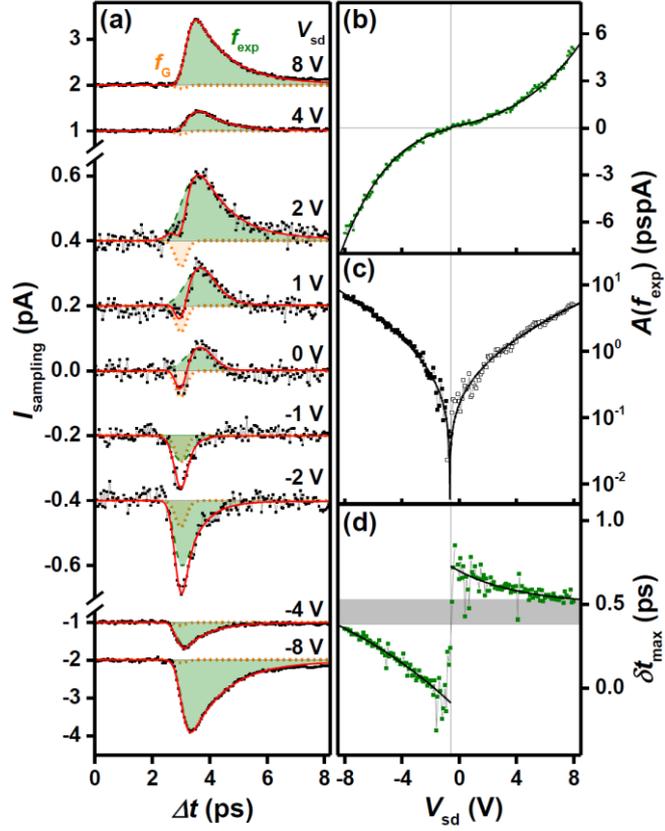

FIG 4. (color online) (a) Time-resolved photocurrent $I_{\text{sampling}}$ vs. $\Delta t$ for different $V_{\text{sd}}$ for CNTs with $d_{\text{CNTS}}$ ~ 1.5 nm. Line is sum of a Gaussian $f_G$ (dotted line) and an exponential decay $f_{\text{exp}}$ (dashed line). (b) Fit area $A(f_{\text{exp}})$ vs. $V_{\text{sd}}$. (c) Logarithmic plot of $A(f_{\text{exp}})$. (d) Relative time delay $\delta t$ between $f_G$ and $f_{\text{exp}}$ as a function of $V_{\text{sd}}$.

# Ballistic and resonant negative photocurrents in semiconducting carbon nanotubes

Christoph Karnetzky,[1,2] Lukas Sponfeldner,[1] Max Engl,[1] and Alexander W. Holleitner[*1,2]

[1]*Walter Schottky Institute and Physics Department, Technical University Munich, Am Coulombwall 4(a), 85748 Garching, Germany.*

[2]*Nanosystems Initiative Munich (NIM), Schellingstr. 4, 80799 Munich, Germany.*

*contact: holleitner@wsi.tum.de

- Supplemental Material -

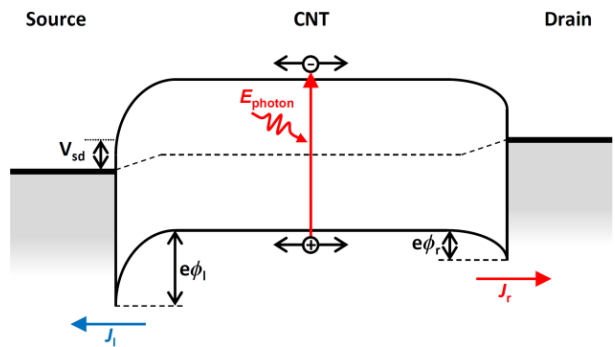

**FIG S1. Schottky emission model.** To describe the photocurrent, we use the Schottky emission model which describes the thermionic emission of charge carriers across a potential barrier that is modified by the applied bias voltage $V_{sd}$ [1][2]. We assume $V_{sd}$ to drop only across the metal-CNT interfaces which is reasonable for small $V_{sd}$ where no dark current passes through the sample (compare regime I in Fig. 2(c) of the main manuscript). For photogenerated electrons in the conduction band, there are no contact barriers. Therefore, the electron currents from the CNTs to the contacts cancel out and do not contribute to the net photocurrent. Hence, we consider only the currents $J_l$ ($J_r$) of photogenerated holes from the CNT to the left source (right drain) contact. The total photocurrent generated in the CNTs can then be expressed as $J_{Schottky} = J_l + J_r = A \cdot \{ \exp[ k_l \cdot \mathrm{Sqrt}(-V_{sd} + V_\Delta) ] - \exp[ k_r \cdot \mathrm{Sqrt}(V_{sd} - V_\Delta) ] \}$, with the amplitude $A$ as the modified Richardson-factor, and $k_{l,r}$ a rate factor depending on the electron temperature and the contact resistivity [1]. In general, the potential barriers at the contacts have different heights. Regarding this, $V_\Delta$ is the voltage that has to be applied for the heights of both barriers to become equal. Hence, for $V_{sd} = V_\Delta$ the photocurrents to both contacts cancel out and $J_{Schottky} = 0$.

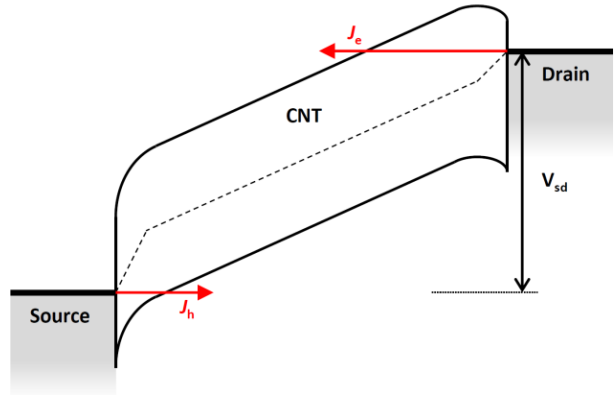

**FIG S2. Fowler-Nordheim tunneling model.** For describing the dark current through the CNTs, we use the Fowler-Nordheim tunneling model which describes the tunneling of charge carriers through a triangular potential barrier [1]. As sketched in Figure S2, the tunnel barriers are generally asymmetric so that we consider in first approximation only the dominating tunnel current of $J_h$ and $J_e$ for the given bias voltage $V_{sd}$. Hence, we model the dark current $I_{dc}$ through the sample with $J_{FN} = A * V_{sd}^2 * \exp[ -b / V_{sd} ]$, with $A$ the amplitude and $b$ a factor depending on the height of the respective contact barrier [1].

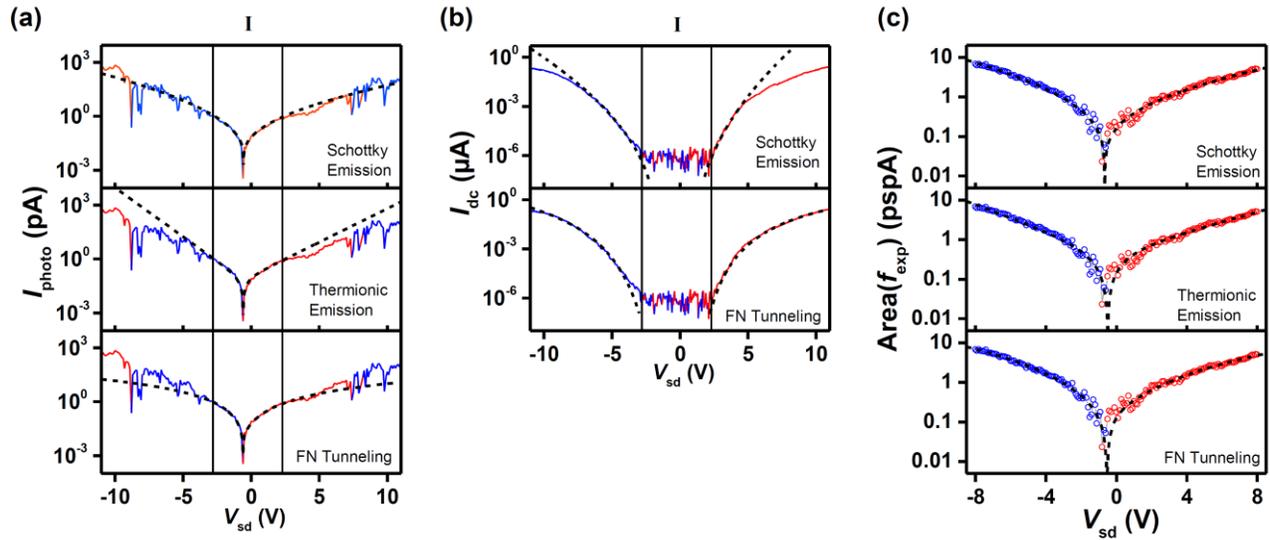

**FIG S3. Comparison of all fit functions.** (a) Logarithmic plots of $I_{photo}$ vs. $V_{sd}$ as in Fig. 2(b) of the main manuscript. Dashed lines are fits according to thermionic emission, Schottky emission, and Fowler-Nordheim (FN) tunneling, respectively. The fit models only use data in regime I where no dark current $I_{dc}$ passes through the sample. The fits are extrapolated for higher voltages. The extrapolation by the Schottky emission model describes the high voltage data best. (b) Logarithmic plots of $I_{dc}$ vs. $V_{sd}$ as in Fig. 2(c) of the main manuscript. The Fowler-Nordheim (FN) tunneling model describes the data best. (c) Logarithmic plots of $A(f_{exp})$ vs. $V_{sd}$ as in Fig. 4(c) of the main manuscript. In principle, all models can reproduce the data reasonably well. This is because the data in the low voltage regime I is too noisy to allow a reliable extrapolation for high voltages. Therefore, the dashed lines are the fits of the respective model to the whole dataset. Generally, we find barrier heights in the order of 30 - 90 meV to describe both the time-averaged and time-resolved photocurrents, which seems to be realistic. However, an exact determination of the barrier height is only possible with a gate-voltage dependence [2]. This is currently impossible with the utilized stripline circuits and the integration of the CNTs via dielectrophoresis [3].

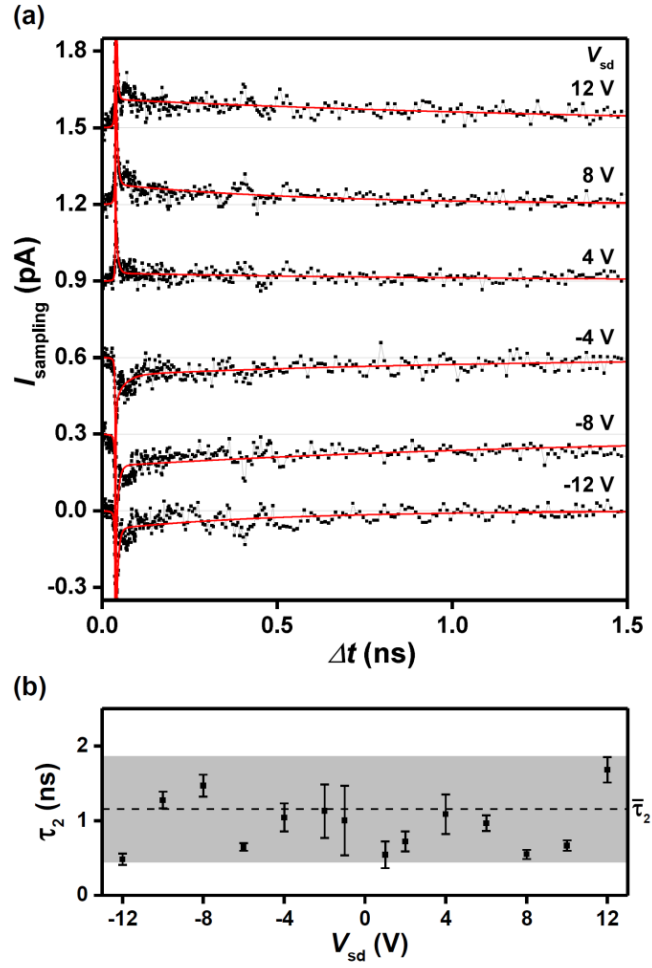

**FIG S4. Lifetime limited photocurrent.** (a) Time-resolved photocurrent $I_{sampling}$ (black dots) vs. $\Delta t$ for the first set of CNTs with $d_{CNTS} \sim 1.5$ nm up to a time delay of 1.5 ns for different $V_{sd}$ with an exponential decay fit (red line). The fit considers the ultrafast processes as discussed in the main manuscript. Moreover, it necessitates a decay time $\tau_2$ in the regime of nanoseconds to describe the data for the long time delays. (b) Decay time $\tau_2$ vs. $V_{sd}$. Gray area depicts confidence level of $\tau_2$.

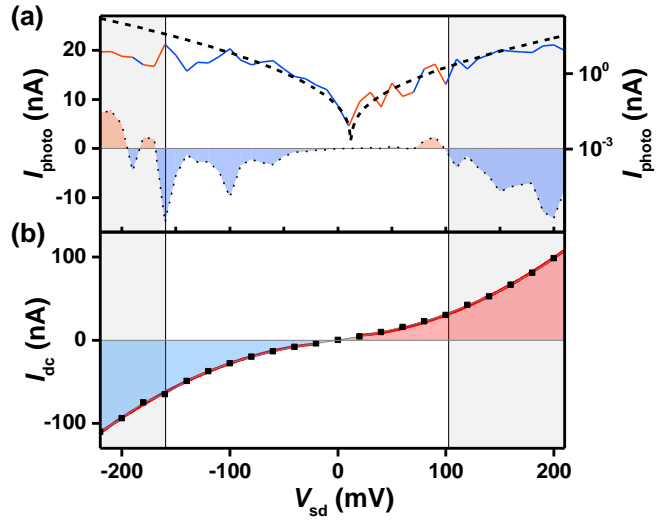

**FIG S5. Time-averaged photocurrent for a CNT of the second set of CNTs (with an excitation of $E_{11}$).** (a) $I_{photo}$ vs. $V_{sd}$ for CNTs with $d_{CNTS}$ ~ 0.8 nm and $E_{photon}$ = 1.23 eV ~ $E_{11}$ in linear and logarithmic scale. Red (blue) color indicates positive (negative) sign of $I_{photo}$. The laser and the experimental parameters are the same as for the results in the main manuscript. The dashed line describes the data according to the Schottky emission model. Again, it is fitted to the data at small bias and interpolated for large biases. (b) Dark current $I_{dc}$ vs. $V_{sd}$ without laser excitation. The red line is the fit according to the Fowler-Nordheim tunneling model as described in Figure S2. As a matter of course, it only works for $V_{sd} \neq 0$.

Most importantly, also for this second set of CNTs, we detect a sign change of $I_{photo}$ when we increase $V_{sd}$, although $I_{dc}$ increases monotonically (gray areas in Figure S5). However, we typically do not identify a regime I where $I_{photo}$ can be measured without $I_{dc}$ passing through the sample. We interpret this finding by the low contact resistance, which we detect for these CNTs which is possibly caused by the specific surface morphology of this second set of CNTs. Hereby, the sign change of $I_{photo}$ in the gray areas of Figure S5 resembles the situation, where an overwhelming part of charge carriers are injected from the contacts, such that a population inversion occurs and $I_{photo}$ changes sign.

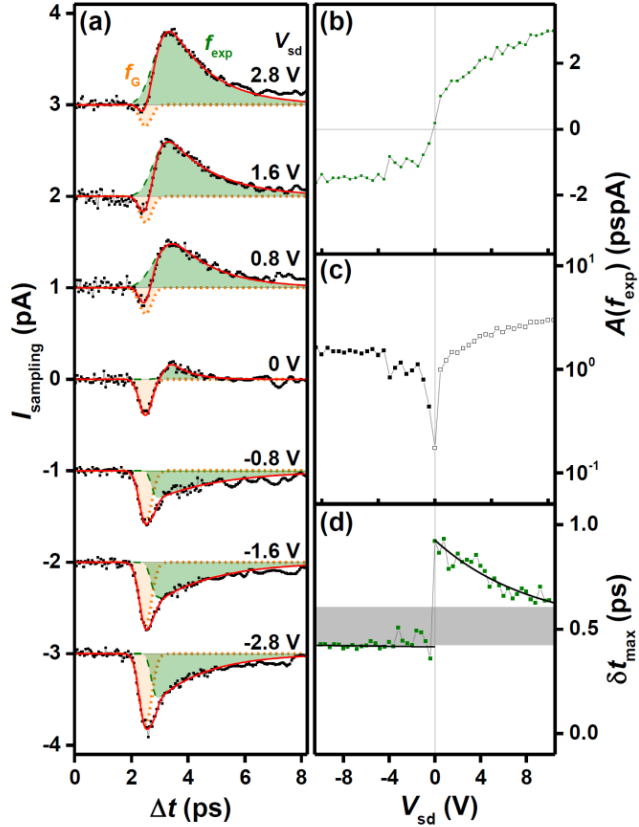

**FIG S6. Time-resolved photocurrent for a CNT of the second set of CNTs (with an excitation of $E_{11}$).** (a) Time-resolved photocurrent $I_{sampling}$ (black dots) vs. $\Delta t$ for different $V_{sd}$ for CNTs with $d_{CNTS}$ ~ 0.8 nm. Line is the sum of a Gaussian $f_G$ (dotted line) and an exponential decay $f_{exp}$ (dashed line). (b) Fit area $A(f_{exp})$ vs. $V_{sd}$. (c) Logarithmic plot of $A(f_{exp})$ vs. $V_{sd}$. (d) Relative time delay $\delta t$ between $f_G$ and $f_{exp}$ as a function of $V_{sd}$. The data are consistent with the CNT having an asymmetric, low contact configuration. In particular, one contact is very small, such that for a negative $V_{sd}$, $\delta t$ is constant ~0.4 ps. In other words, also for a small $V_{sd}$, the voltages drops across the whole CNT. For a positive $V_{sd}$ approaching zero bias, $\delta t$ reaches (0.8 ± 0.1) ps. Both values are consistent with a ballistic transport of the photogenerated charge carriers. Moreover, the data again can be interpreted in a way that photogenerated excitons are dissociated at large fields, i.e. at one contact and at a large bias, as discussed in the main manuscript for the first set of CNTs.